# Strain induced pressure effect in pulsed laser deposited thin films of the strongly correlated oxide $V_2O_3$

S. Autier-Laurent[1], B. Mercey[1], D. Chippaux[1], P. Limelette[1,2], Ch. Simon[1]

[1]Laboratoire CRISMAT, CNRS UMR 6508, ENSICAEN, 6 bd du Maréchal Juin, F-14050 Caen cedex, France

[2]Laboratoire LEMA, CNRS-CEA UMR6157, Université F. Rabelais, Parc de Grandmont, 37200 Tours, France

$V_2O_3$ thin films about 1000Å thick were grown on $Al_2O_3$ (0001) by pulsed laser deposition. The XRD analysis is in agreement with R-3c space group. Some of them exhibit the metal / insulator transition characteristic of $V_2O_3$ bulk material and others samples exhibit a metallic behavior. For the latter, the XPS analysis indicates an oxidation state of +III for vanadium. There is no metal / insulator transition around 150 K in this sample and a strongly correlated Fermi liquid $\rho \sim AT^2$ behavior of the resistivity at low temperature is observed, with a value of A of $1.2\ 10^{-4}$ $\Omega$ cm, 3 times larger than the bulk value at 25 kbar.

**Introduction**

Vanadium oxide $V_2O_3$ is considered as the prototype of the metal / insulator transition (MIT) and is a very active field of both experimental[1,2] and theoretical[3,4,5] physics. It is only very recently that thin films of $V_2O_3$ were obtained[6], showing that the MIT can be suppressed in very thin films. So, we have synthesized $V_2O_3$ films in order to investigate the strain effect induced by the substrate, that can act as an external pressure, and the strength of the electronic correlations in the obtained metallic state.

At room temperature, $V_2O_3$ has a corundum structure with trigonal symmetry, and behaves as a rather good metal, with a room temperature resistivity of the order of $10^{-3}$ $\Omega$ cm. When the temperature is decreased below about 150 K, it undergoes a transition from a paramagnetic metallic phase to an antiferromagnetic insulating phase, marked by an increase in resistivity of about seven orders of magnitude and accompanied by a large change in crystal structure from trigonal (corundum–type) to monoclinic. The MIT in $V_2O_3$ sensitively depends on the externally applied pressure, on the doping and on the oxygen stoichiometry[7]. This suggests that the MIT is not only affected by the external pressure but also by substrate strain-induced pressure in the films. Previous studies, carried out in our laboratory, have already shown that the substrate-induced strain promoted / initiated large changes in the magnetic and electronic transport behavior of the charge-ordered and orbital-ordered manganites[8]. Also, it is interesting to study the electronic transport properties associated to the influence of the MIT in $V_2O_3$ deposited on a substrate applying a biaxial in-plane strain to the $V_2O_3$ structure. In a previous study[9], it was shown that it is possible to grown in epitaxy such a film, but no determination of the oxidation state of the vanadium was performed.

**Experimental**

$V_2O_3$ thin films were grown by pulsed laser deposition (PLD) using a multi-target computer controlled deposition system. This system differs from the others systems used in the laboratory, since the laser beam, reflected by a motorized computer-controlled mirror, sweeps the surface of the target. In the others systems, to avoid preferential erosion, the target continuously rotate. The PLD method requires the use of a ceramic target. Since it is difficult to obtain a large enough amount of $V_2O_3$ to prepare a highly densified target usable for the PLD, two routes could be used for the growth of the $V_2O_3$ films with the stabilization of the +III oxidation state of the vanadium: either a metal target associated with a precise control of the oxidation or a $V_2O_5$ target associated with a fine control of the reduction. To determine the stability region of the various oxidation states of the vanadium, the Ellingham diagram was calculated from the free energy $\Delta^r G°$. The variations of $\Delta^r G°$ versus temperature shows that the two possible routes can be used, the +III oxidation state of the vanadium being stabilized either by oxidation of the vanadium or by reduction of the oxide $V_2O_5$. The use of a $V_2O_5$ target was decided for two reasons. Firstly, the possible over-heating of the metal target might lead to a large amount of droplets. Secondly, it was elegant to take benefit of the PLD method, which is a reducing process when no oxygen is intentionally added to the atmosphere. The formation of the vanadium +III is then controlled by the sticking coefficients of the vanadium and of the oxygen, provided by the target, on the substrate at the deposition temperature. As the use of a metallic target for a PLD growth might lead to a large concentration of droplets resulting from an over-heating of the metallic target, the use of a $V_2O_5$ oxide target was decided. The sintered high-density $V_2O_5$ target was prepared in the laboratory.

For the PLD experiments, a Lambda Physik KrF excimer laser (wavelength: 248nm) was focused on the target with a repetition rate of 3Hz. Films were grown on [0001]-oriented sapphire. This substrate presents a trigonal structure whose lattice parameters, described in the hexagonal setting are: *a = 4.75 Å* and *c = 12.99 Å*. As bulk material $V_2O_3$ crystallizes in a trigonal cell corundum-type, this structure can be described from $MO_6$ octahedrons that shear 3 of their edges to form a succession of $MO_3$ layers providing the $M_2O_3$ stoichiometry. The space group of this compound is R-3c and the cell parameters are, in the hexagonal setting, *a = 4.9515 Å* and *c = 14.003 Å*. Assuming an epitaxial growth of the $V_2O_3$ films on the sapphire substrate the in-plane lattice mismatch between the film and the substrate is found to be of the order of – 4 %. Since there is a rather large in-plane biaxial compression, one could thus expect a significant extension of the c-axis in the films.

The structural study of the films has been investigated with a θ-2θ X-ray diffraction (XRD), using a Seifert XRD 3000P (Cu $K_{\alpha 1}$ radiation, λ = 0.15405 nm), for the out of plane lattice parameter *c* and with a four circle diffractometer for the in-plane lattice parameters. In agreement with the space group R-3c of the bulk material, only the *00l* reflections on the XRD pattern can be observed with l = 6n. No other phase or orientation was detected in the films. The study of the in plane lattice parameters showed that the films are epitaxially grown on the sapphire substrate.

To determine the oxidation state of the vanadium, in the $V_2O_3$ films, X-ray photoelectron spectroscopy (XPS) has been carried out in an analytical system using a Leybold hemispherical analyzer. The spectrometer calibrated with Cu, Ag, and Au test samples has a relative resolution of about 1% and was used at a pass energy of 100 eV except for the core levels V2p and O1s recorded at 50 eV. The excitation was provided to the sample using a non-monochromatized Mg X-ray source (Mg K$_\alpha$, E$_{h\nu}$=1253.6 eV). The films were analyzed without ion etching to avoid any modification of the vanadium oxidation state. The Al 2p line from the low conductivity substrate was used to correct for charging effects that lead to shifts in line positions. This correction might be of importance since the binding energies of the different oxidation states of the vanadium are pretty close (table 1). All experimental spectra are Shirley background corrected. The satellites are also subtracted for the XPS spectra. The fit of recorded regions were supplied by the Leybold data system which uses an intermediate lorentzian / Gaussian (L/G) function, first introduced by Frazer and Suzuki[10] and defined as follows:

$$I(E) = \frac{I_0}{\left(1 + 4 \times \left(2^{\left(\frac{1}{n}\right)} - 1\right) \times \left(\frac{E - E_0}{FWHM}\right)^2\right)^n}$$

L/G mixing ratio: n=1 Lorentzian, n $\rightarrow \infty$ Gaussian
Maximum: $I_0, E_0$
Full width at half maximum: FWHM

The transport properties of the films have been investigated as a function of temperature in the range 10 K-360 K using the four-point method with a Physical Properties Measurements System (PPMS) from Quantum Design. Electrical contacts were made with four silver pads thermally evaporated on the film surface through a mask, and then connected to the measuring system by Al / Si wires welded by ultrasounds using a West Bond 7674D bonding machine.

**Results**

Using a V$_2$O$_5$ target, the thin films grow with a substrate temperature between 600°C and 740°C, according to the stability domain of V$_2$O$_3$. To get a good thermal contact between the heater and the substrate, a silver paste is used to glue the substrate to the heater. To reduce the oxygen concentration in the film with respect to the target, the growth is carried out in a pure argon atmosphere in a pressure ranging from 0.20 mbar to 0.02 mbar. Then films about 1000 Å thick are deposited in these conditions, after the growth the films are cooled in the same atmosphere than during the deposition or under vacuum. To obtain reproducible deposition conditions and particularly to control the oxygen content in the chamber prior to the deposition and before introducing the argon, the pressure of the system is stabilized at 10$^{-6}$ mbar (base pressure of the system 10$^{-8}$ mbar), with the heater at the deposition temperature.

First, the argon pressure is 0.2 mbar and the temperature is varied to obtain the best crystallization of V$_2$O$_3$. So, one finds that the films grown at a temperature higher than 700°C are badly crystallized with several crystallographic orientations as attested by their X-ray diffraction diagram. The films grown at 600°C exhibit only one reflection with a low intensity. Only the X-ray diffraction diagrams of the films grown at 650°C display two weak reflections in agreement with the expected values for a V$_2$O$_3$ film. In a second step, when the argon pressure is lowered down to 0.02 mbar, a better crystallization of the films is reached, characterized by the two expected sharp reflections *006* and *0012* in the θ-2θ scan. Once the growth conditions giving the best XRD has been determined, the transport properties have been measured as a function of temperature from 300 K down to 5 K. As shown in the Figure 1, three behaviors can be distinguished according to their transport properties and their ***c*** lattice parameters. In this figure, all the films were grown under the same temperature and pressure conditions (600°C and 0.02 mbar). Only the cooling procedure is different.

First, let us remind that the bulk material undergoes a transition from a high temperature paramagnetic metal to a low temperature antiferromagnetic insulator with a discontinuous increase of the resistivity of about seven orders of magnitude below 150 K[7]. Thin films of type 1 (cooling down in vacuum), which have a *c* lattice parameter smaller than the bulk one, have a similar behavior with the bulk (MIT around 150 K).

Strikingly, for the type 2 (cooling down in argon atmosphere 0.02 mbar) this MIT is completely suppressed and films are metallic at low temperature as clearly demonstrated in the Figure 1 and observed for thinner films by Qiang Luo *et al.*[6]. Thin films of type 3 (cooling down in argon atmosphere 0.2 mbar) are also metallic down to the low temperatures but they present an incomplete metal to insulator transition around 150 K. The two types of films which are the most metallic (type 2 and 3) exhibit a *c* lattice parameter larger than the bulk. Taking into account the previous work concerning the manganites thin films[8], the absence of the MIT in our films deposited on sapphire could originate from the substrate-induced strain effect. Indeed, while the MIT in bulk material is associated with a large structural change from trigonal to monoclinic, the atomic positions in the films are constrained by the substrate ones. Thus, no large structural change can occur and the MIT becomes unfavorable.

Typically, for thin films of type 2 the inter-reticular distances of the reflections *006* and *0012*, respectively 2.34 Å and 1.17 Å, linked to a *c* lattice parameter of 14.04 Å are in a good agreement with the bulk one along the [001] direction (*c = 14.003 Å*). Assuming a strong in-plane compression and according to the Poisson's law, the out of plane parameter should be larger. Considering the volume of the cell of the bulk material V = 297 Å$^3$ and assuming an in-plane lattice parameters of the film equal to these of the substrate, an out-of-plane lattice parameter of 15.16 Å is expected, namely larger than in the bulk. The volume of the cell of the film is markedly different from the bulk one. This result differs from those obtained by different authors who measure out-of-plane lattice parameter even smaller than the bulk one[11,12] for films deposited by thermal evaporation or reactive electron-beam evaporation on sapphire substrates. The values of the out-of-plane lattice parameters for our films having the same thickness and deposited in the same conditions are between 14.02 and 14.10 Å (type 2 and 3). Using a four circle X-ray diffractometer we have measured in a film with an out-of-plane lattice parameter of 14.10 Å an in-plane lattice parameter equal to 4.89 Å, namely slightly smaller than the bulk one. So, the volume of the unit cell of the film is now $V_{film}$ = 292 Å$^3$, while in the bulk material this volume is $V_{bulk}$ = 297 Å$^3$. The measured volume of the unit cell of the film is in agreement with the expected value: an in-plane compression induces a small decrease of the in-plane parameters and an increase of the out-of plane parameter with respect to the bulk one.

To precisely determine the oxidation state of the vanadium in films of type 2 which do not behave as bulk $V_2O_3$ material, a XPS study is carried out. A typical spectrum of the vanadium and oxygen lines is shown in fig. 2a. The strongest line attributed to the O 1s level can be fitted as two peaks centered at 529.8 eV (FWHM = 1.8 eV, L/G mixing ratio = 2) and 531.5 eV that correspond respectively to the V-O bound and to the physically adsorbed oxygen at the surface of the film. Due to spin orbit coupling the V 2p level splitting into two lines V $2p_{1/2}$ and V $2p_{3/2}$ is found to be 7.6 eV with a binding energy of the major line V $2p_{3/2}$ equal to 515.5 eV (FWHM = 4.1 eV, L/G mixing ratio = 4). Despite the presence of surface contamination, we obtain a well-defined XPS spectrum. The separation between the bulk and the surface signal for oxygen shows a minor pollution which exhibits no typical structure in the vanadium $2p_{3/2}$ spectrum which is in good agreement with previous works[13,14,15,16,17].

In the valence band region, figure 2b, according to the $V_2O_3$ literature[13,14], two peaks are observed: a low binding energy peak at about 0.8 eV from the Fermi edge due to the V3d electrons not involved in the V-O bonding, and a broad line at about 5 eV attributed to O2p / V3d hybrid states. It is worthy to note that the $V_2O_5$ compound has no 3d band and $VO_2$ just a weak one[13].

The Auger spectrum of the major lines initiated by the O1s and V2p photoemission is also of great interest because its profile is different from $V_2O_3$ to $VO_2$ and $V_2O_5$. Compared to the work of Sawatzky et al.[13], our Auger spectrum is well identified as the $V_2O_3$ one (fig. 2c). Analysis of the results mentioned above, in agreement with previous XPS studies on bulk and thin film of $V_2O_3$ [13,14,15,16,17] denotes an oxidization compatible with $V^{3+}$ state and shows that our thin films of type 2 are well identified as $V_2O_3$ compound.

As one observes in the fig. 1, the films of type 2 exhibit a metallic resistivity over the investigated range temperature. The rather low value of the residual resistivity attests of the low density of grains boundaries, in agreement with an epitaxial growth of the films. Moreover, it results that the temperature dependences of the resistivity reported in fig. 1 are qualitatively similar to the measured ones in single crystals at high pressures[18]. In particular, the Fermi liquid regime is recovered at low temperatures with a resistivity varying as $\rho = \rho_0 + AT^2$, $\rho_0$ being the residual resistivity. By plotting in the Fig. 3 resistivity as a function of $T^2$, this Fermi liquid regime can be compared with those determined in single crystals at high pressures[18]. Interestingly, one deduces a Fermi liquid transport parameter A in thin films higher than in crystals, with actually A~1.2 $10^{-4}$ m$\Omega$ cm K$^{-2}$.

Since the parameter A is expected to be proportional to the square of the electronic effective mass as A$\propto$(m*)², this result seems to suggest a significant enhancement of m* in the film compared with the single crystals. The effective mass being representative of the strength of the interactions between electrons, one can also infer that the metallic regime observed in the film is higher correlated than in the crystals under the considered pressures.

**Conclusion**

Thin films of $V_2O_3$ were grown by pulsed laser deposition from a $V_2O_5$ sintered target using reducing conditions during the growth. XRD measurements confirm the structure of $V_2O_3$. According to their transport properties three types of samples were determined. The first one presents the same *c* lattice parameter than the bulk form and exhibit a metal to insulator transition around 150 K. The two others types of samples are very interesting since they are metallic at low temperature. XPS analysis, carried out on the metallic films, is in agreement with a +III oxidation state for vanadium. The unit cell is markedly modified (this might induce an important change in the interactions since the V-O distances are modified by this strain). This regime of strain deeply modifies the transport properties by stabilizing in particular the metal down to the low temperatures. Since the metal-insulator transition around 150 K in single crystals can be suppressed in some of the films, further experimental investigations should be for now considered in order to characterize this low temperature strongly correlated metallic regime.

*S. Autier-Laurent acknowledges the «Région de Basse Normandie » for its financial support.*

# FIGURE CAPTIONS

Figure 1 : Resistivity versus temperature of the 3 types of films grown by PLD

Figure 2: a) XPS spectrum of V2p and O1s core levels of $V_2O_3$ thin film of type 2

b) Valence- band region of the XPS spectra. Type-2 films compared to the study of Sawatzky et *al* [13]

c) O and V Auger lines of type-2 films compared to the work of Sawatzky et *al*[13]

Figure 3 : Resistivity versus $T^2$. A comparison is made with the results obtained for single crystals under hydrostatic pressure[18].

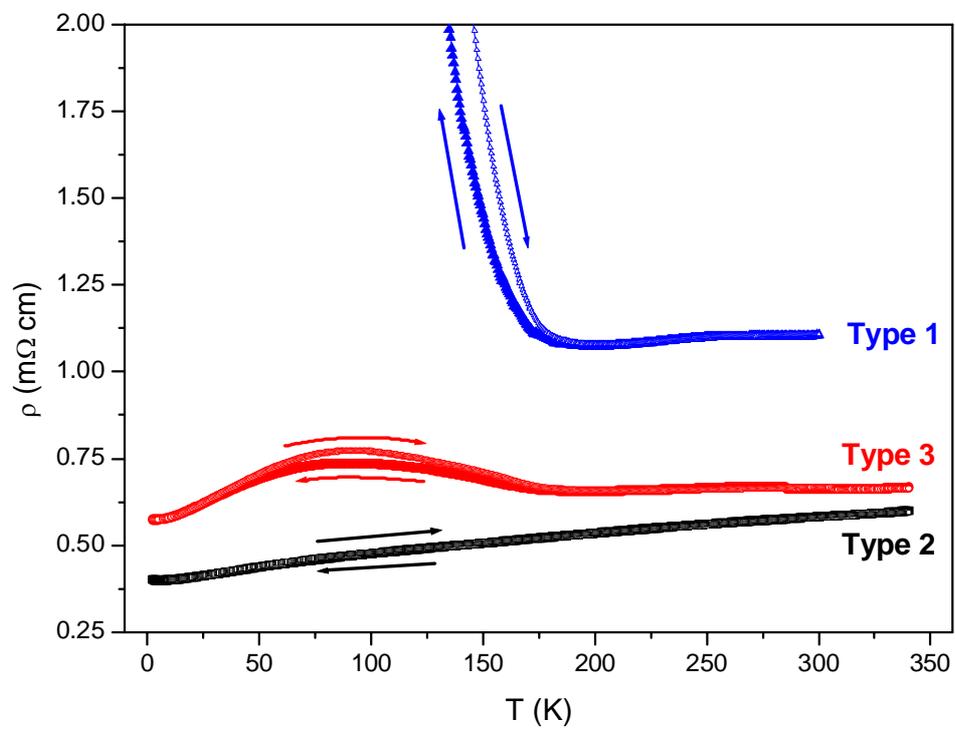

Figure 1

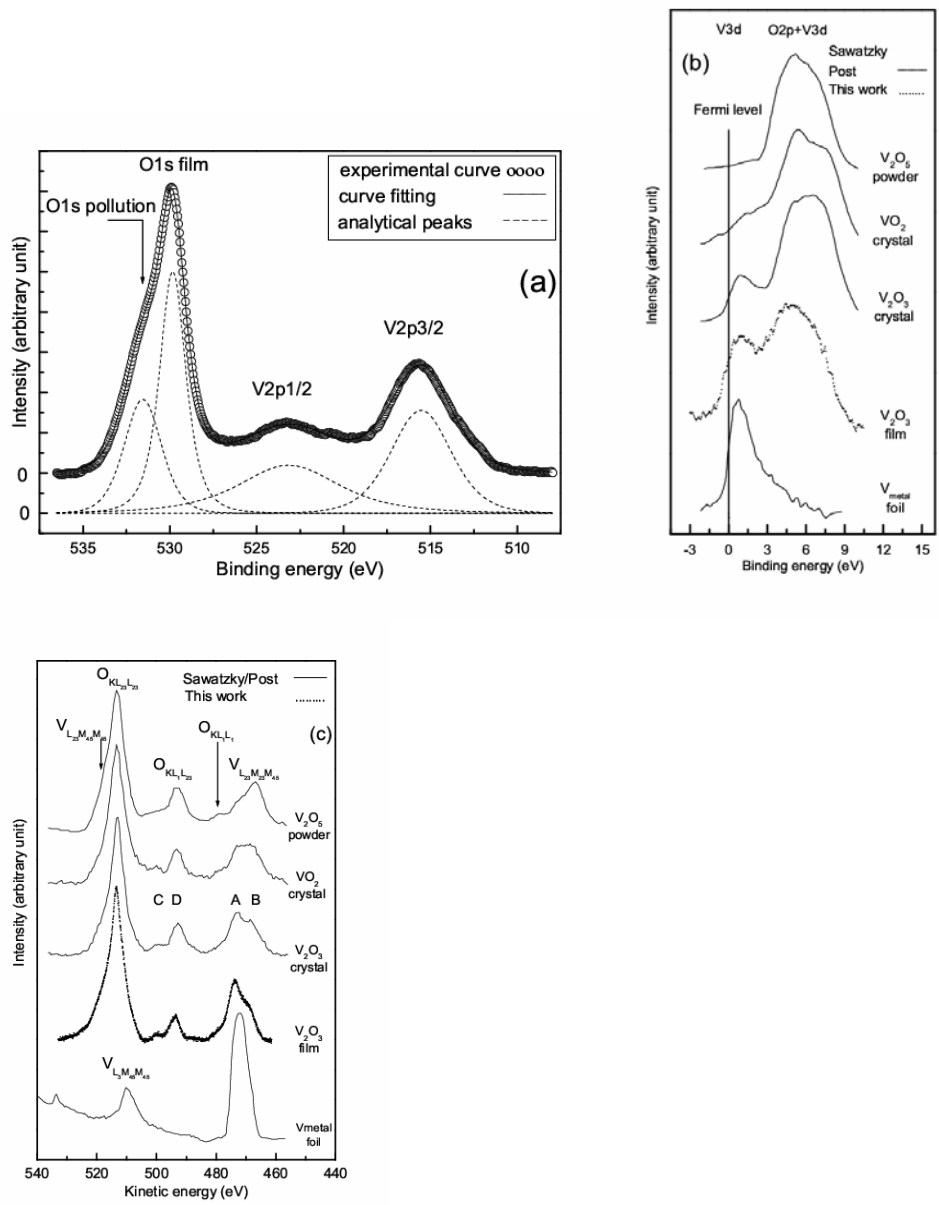

Figure 2

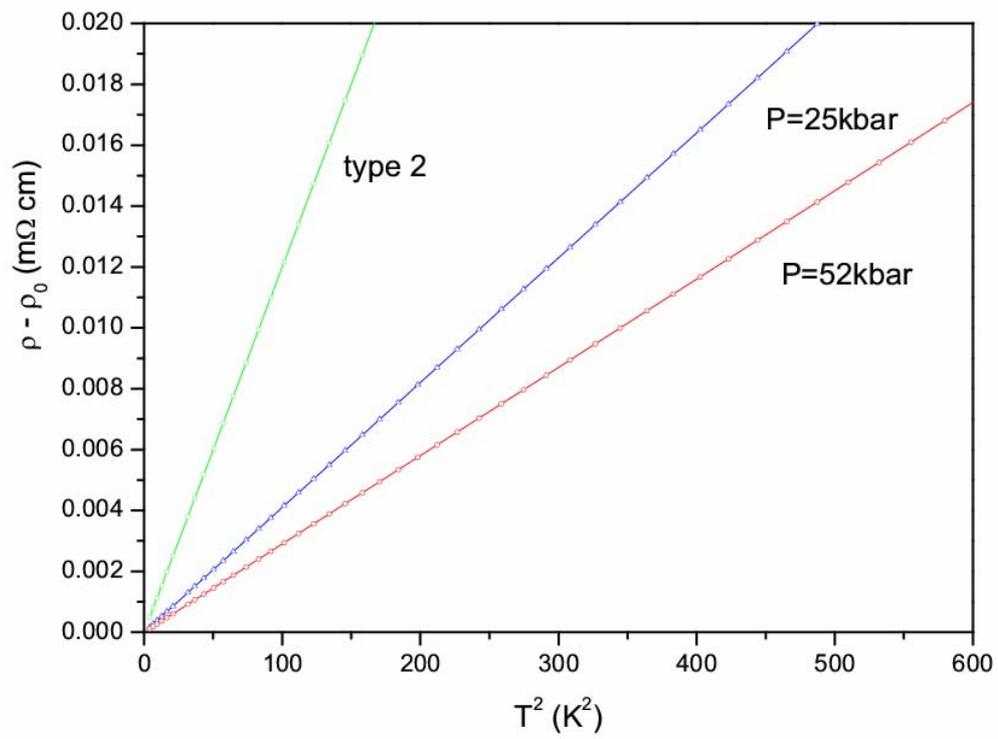

Figure 3

[1] P. Limelette, A. Georges, P. Wzietek, D. Jerome, P. Metcalf, J.M. Honig, Sciences, **302**, 89 (2003)

[2] M. Yethiraj, J. Solid State Chem., **88**, 53 (1990)

[3] G. Kotliar, Science, **302**, 67 (2003)

[4] N.F. Mott, *Metal-Insulator transitions*, Taylor and Francis, London (1974)

[5] H.K. Kim, H. You, R.P. Chiarello, H.L.M. Chang, T.J. Zhang and D.J. Lam, Phys. Rev. B, **47**, 12900 (1993)

[6] Qiang Luo, Qinin Guo and E.G. Wang, Appl. Phys. Lett., **84**, 2337 (2004)

[7] D.B. Mc Whan and J.P. Remeika, Phys. Rev. B, **2**, 3734 (1970)

[8] W Prellier, A.M. Haghiri-Gosnet, B. Mercey, Ph. Lecoeur, M. Hervieu, Ch. Simon and B. Raveau, Appl. Phys. Lett. **77**, 1023 (2000)

[9] S. Yonezawa, Y. Muraoka, Y. Ueda, Z. Hiroi, solid state comm. 129, 245 (2004).

[10] R. D. B. Frazer, E. Suzuki, Anal. Chem. 41, 37 (1969)

[11] I. Yamaguchi, T. Manabe, T. Kumagai, W. Kondo, S. Mizuta, Thin Solid Films, **366**, 294-301 (2000)

[12] H. Schuler, S. Klimm, G. Weissmann, C. Renner, S. Horn, Thin Solid Films, **299**, 119 (1997)

[13] G.A. Sawatzky and D. Post, phys. Rev. B, **20**, 1546 (1979)

[14] A.C. Dupuis, M. Abu Haija, B. Richter, H. Kuhlenbeck, H.-J. Freund, Surface Science **539**, 99 (2003)

[15] J. Mendialdua, R. Casanova, Y. Barbaux, J. Electron Spectrosc. Relat. Phenom., **71**, 249 (1995)

[16] Q. Guo, D.Y. Kim, S.C. Street, and D.W. Goodman, J. Vac. Sci. Technol. **A 17(4)**, 1887 (1998)

[17] M. Demeter, M. Neumann, W. Reichelt, Surface Science **454**, 41 (2000)

[18] D.B. Mac Whan and T.M. Rice, Phys. Rev. Lett., 22, 887 (1969)